\documentclass{amsart}

\usepackage{graphicx}

\usepackage{amssymb} 

\newtheorem{theorem}{Theorem}[section]

\theoremstyle{definition}

\theoremstyle{remark}

\numberwithin{equation}{section}

\usepackage{eqnarray,amsmath}

\newcommand{\R}{\mathbb{R}}

\renewcommand{\S}{\mathbb{S}}

\newcommand{\SL}{SL_2(\R)}
\newcommand{\USL}{\widetilde{\SL}}

\newcommand\norm[1]{\left\lVert#1\right\rVert}

\newcommand{\X}[1]{\frac{\partial}{\partial x_{#1}}}

\renewcommand{\dot}[2]{\langle#1,#2\rangle}

\begin{document}

\title{Design and visualization of Riemannian metrics}

\author{Tiago Novello}
\address{VISGRAF Laboratory,Rio de Janeiro,Brazil}
\email{tiago.novello90@gmail.com}
\author{Vinicius da Silva}
\address{VISGRAF Laboratory,Rio de Janeiro,Brazil}
\email{dsilva.vinicius@gmail.com}
\author{Luiz Velho}
\address{VISGRAF Laboratory,Rio de Janeiro,Brazil}
\email{lvelho@impa.br}

%    General info
%\subjclass[2000]{Primary 54C40, 14E20; Secondary 46E25, 20C20}

\date{May 7, 2020}% and, in revised form, June 22, 2001.}


\keywords{Riemannian geometry, Graph of a function, Space deformation.}

\begin{abstract}
Local and global illumination were recently defined in Riemannian manifolds to visualize classical Non-Euclidean spaces. This work focuses on Riemannian metric construction in $\R^3 $ to explore special effects like warping, mirages, and deformations.
We investigate the possibility of using graphs of functions and diffeomorphism to produce such effects. For these, their Riemannian metrics and geodesics derivations are provided, and ways of accumulating such metrics. We visualize, in ``real-time'', the resulting Riemannian manifolds using a ray tracing implemented on top of Nvidia RTX GPUs.
\end{abstract}

\maketitle

\begin{figure}[hh]
    \centering
    \begin{tabular}{ccc}
        \includegraphics[width=0.31\columnwidth]{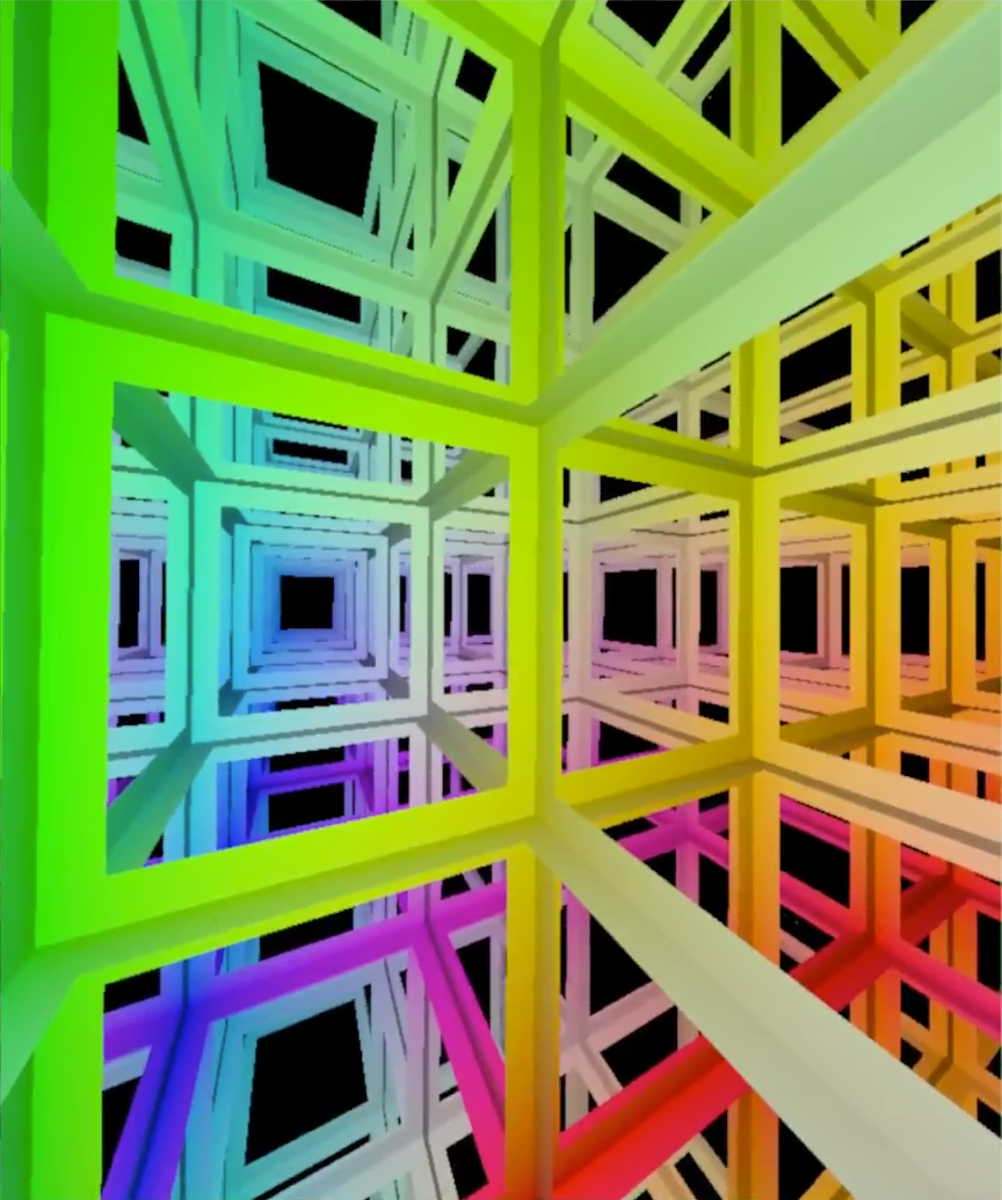}
        &
        \includegraphics[width=0.31\columnwidth]{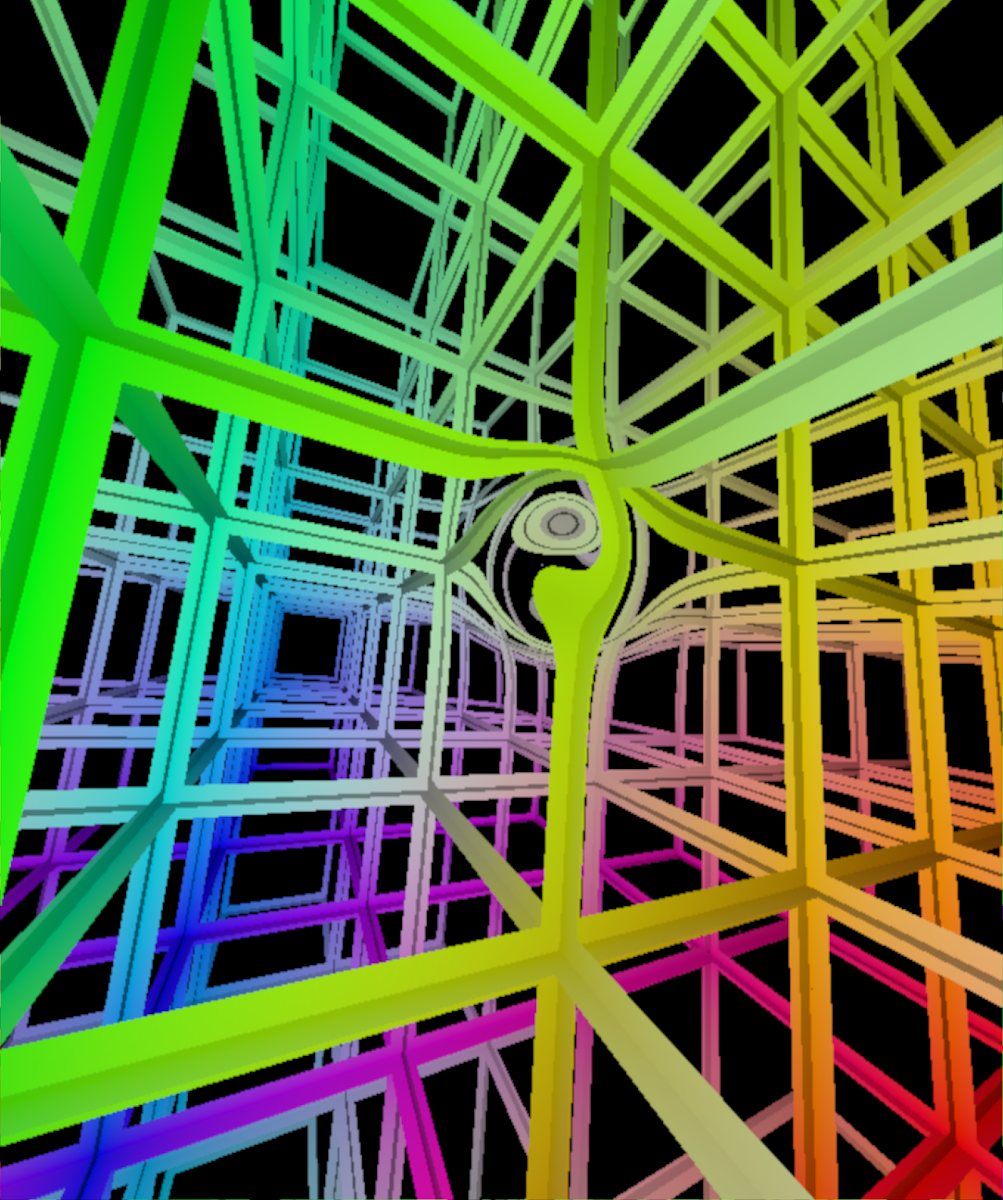}
        &
        \includegraphics[width=0.31\columnwidth]{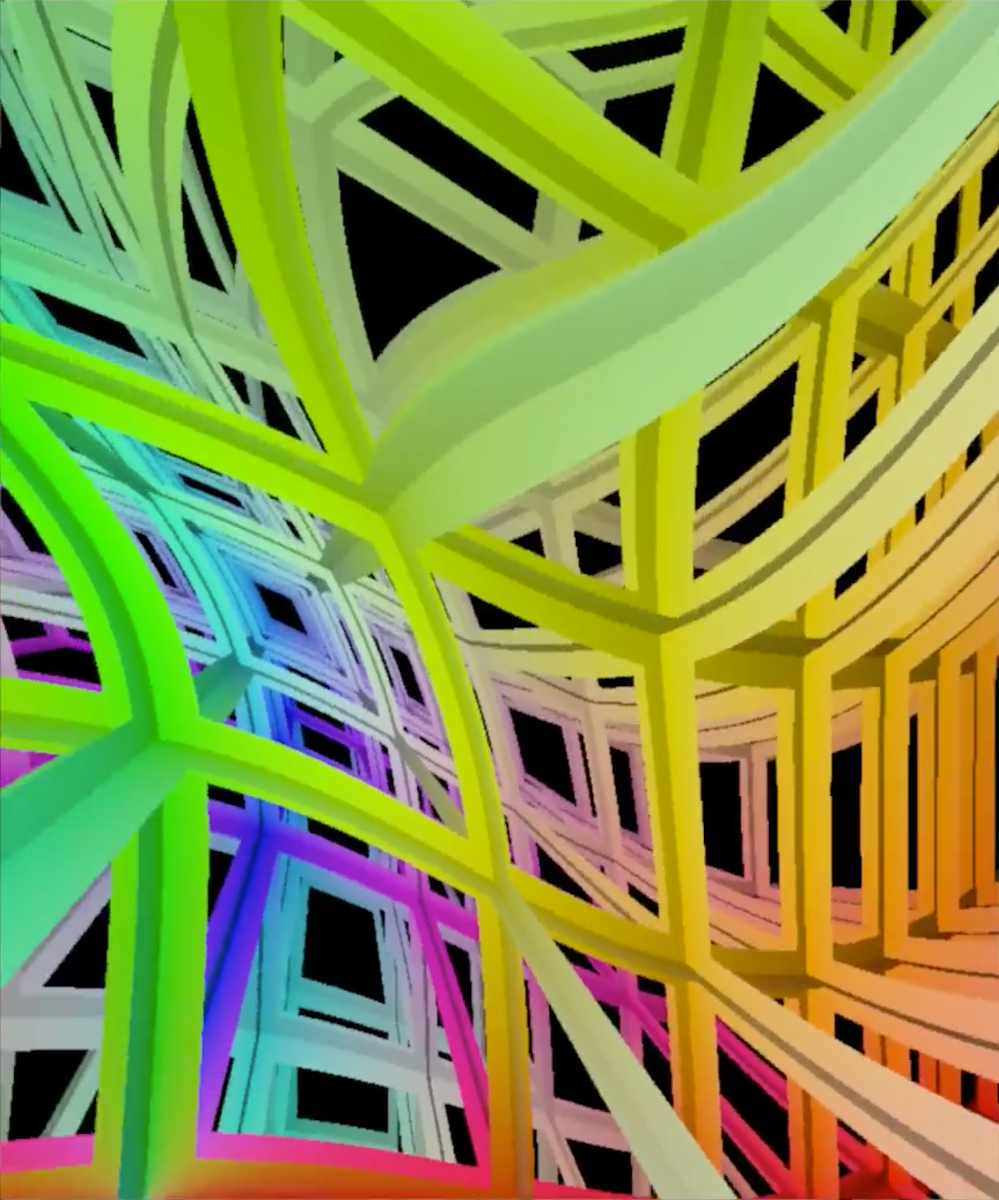}
    \end{tabular}
    \label{fig:six_flat_manifolds}
\end{figure}
%===================================================

%===================================================
\section{Introduction}\label{s-introduction}
%===================================================
% . Ray tracing
\textit{Ray tracing} is a set of techniques used in computer graphics to render photorealistic inside views of a scene in the Euclidean space~\cite{whitted1979improved}. It considers that the light propagates along with straight lines. To ray trace a scene, rays are launched from each image pixel: where the light arrives. If these \textit{primary rays} intersect some surface, the color is computed using \textit{direct} and \textit{indirect} light contributions. The direct illumination considers the rays coming directly from light sources and the indirect gathers the light rays bounced by other surfaces: \textit{secondary} rays.

Ray tracing works intrinsically in space geometry since light travels along lines minimizing lengths (rays). Thus, unlike in rasterization, no tricks to compute information based on the global physical behavior of light are required. In particular, ray tracing could be extended to not necessarily be restricted to Euclidean geometry tracing straight lines. We believe Riemannian geometry is the right setting for this extension since it generalizes the required concepts: metric and rays.

Riemannian space construction has topology as a global geometric constraint. In \cite{velho2020immersive}, the authors define Riemannian shading to explore Nvidia RTX GPUs to visualize Non-Euclidean spaces endowed with non-trivial topology (dating back to Thurston's geometrization conjecture). Instead, this work firstly focuses on time/user-dependent Riemannian metric constructions on the classic space to explore special effects like warping, mirages~\cite{stam1996ray} and scene deformation \cite{barr1986ray}. This implies spaces with trivial topology, but with generic Riemannian geometry. Then, we define Riemannian ray tracing to visualize scenes endowed with such effects. Our first results use graphs of functions and diffeomorphisms to construct the metrics, allowing the modeling of expressive effects. Finally, gaussian bump functions restrict these constructions providing local deformations~\cite{riemannianrayvr}.

That approach opens many geometric questions concerning new ways of representing rays and space. Thus, the goal of this paper is to establish the basis of a new line of research based on employing Riemannian geometry in ray tracing techniques. We believe curved rays can advance the state of the art in many areas, not restricted to rendering only. 
%===================================================

%===================================================
\section{Related works} \label{s-related-works}
%=================================================
In the literature, an expressive number of extensions and modifications of ray tracing were studied to increase efficiency, photorealism, functionality, among others. Almost all of these works focus on tracing straight line rays. Only a few approaches use nonlinear rays; we relate some of these.

% . Alan Bar | Deformations
Bar~\cite{barr1986ray} used a nonlinear approach to ray trace smooth surfaces efficiently. The surfaces are considered to be bent flat planes, which are generated by a deformation of the space. Such deformation corresponds to a coordinate system, where the intersection problem considers the surface being flat, and the rays being curved.

% . Jos Stam |Mirages
In \textit{visualization of physics}, Stam and Languénou~\cite{stam1996ray} presented a ray tracing algorithm integrating perturbations with the basic equations from geometrical optics that govern the propagation of rays in a medium varying its index of refraction continuously. The curvature of the rays depends on the air's index of refraction.
% . Non-linear raytracing
Nonlinear ray tracing was also considered~\cite{groller1995nonlinear} in visualizing relativistic effects, the geometric behavior of nonlinear dynamical systems, and the movement of charged particles in a force field (e.g., electron movement).

% . Our works on Non-Euclidean geometry
In \textit{visualization of mathematics}, ray tracing appeared recently in the visualization of Thurston geometries~\cite{vc-rtorb-2014, velho2020immersive, nilsolsl2}. Berger et al.~\cite{vc-rtorb-2014} used ray tracing to visualize Non-Euclidean spaces. However, this work does not provide real-time immersive and interactive visualization, and it was restricted to Euclidean and hyperbolic manifolds, where rays are modeled by straight lines. 

Velho et. al~\cite{velho2020immersive, nilsolsl2} introduced a framework, implemented on top of Nvidia RTX GPUs, for real-time immersive and interactive visualization of Euclidean, Hyperbolic, Spherical, Nil, Sol, and $\USL$ geometries: the six more interesting Thurston geometries. They introduced a novel ray tracing model for Riemannian manifolds focusing primarily on the topology of the underlying manifolds rather than the geometry. In this work, we consider only space $\R^3$ which has a trivial topology, however, we endow it with complicated and interesting geometries.
%===================================================

%===================================================
% \section{Core concepts} \label{s-core-concepts}
% %===================================================
% . Parameterization (imagem mostrando o R2 parameterizando uma superfície)

% . Riemannian metrics on $\R^3$

% . Tracing curved rays

% . Local deformation using Riemannian geometry

% . Efficient storing of metrics to compute geodesics
%===================================================

%===================================================
\section{Riemannian geometry}\label{s-geometry}
%===================================================
Riemannian geometry~\cite{carmo1992riemannian} was developed from the study of surfaces in $\R^3$. There is a natural metric in each tangent plane of a surface $S\subset \R^3$ inherited from the scalar product of $\R^3$. This implies a distance measure in $S$. \textit{Geodesics} are curves in $S$ minimizing distance. We use the terms \textit{geodesic} and  \textit{ray} interchangeably. 

This section presents the definitions for Riemannian metrics in $\R^3$ and their geodesics. Our ray tracing will operate in this setting. We follow a simplified version of the exposition in \cite{nilsolsl2} which was inspired in the notation of Carmo~\cite{carmo1992riemannian}.

%===================================================
\subsection{Parametric manifold}\label{subsec:space}
\begin{figure}[ht]
	\centering
	\includegraphics[scale=0.29]{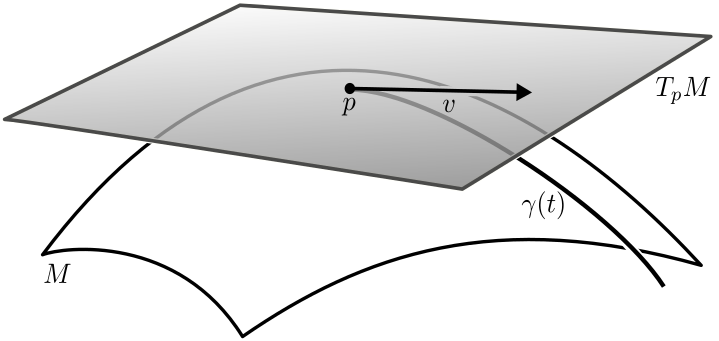}
	\vspace{-0.2cm}
	\caption{A $2$-manifold $M$ and its tangent space $T_pM$ at the point $p$. The geodesic $\gamma(t)$ starts at $p$ in the direction $v$.}
	\label{fig:geodesic}
\end{figure}

\hfill\\ 
A \textit{parametric $3$-manifold} $M$ is a topological space which can be parameterized by $\mathbb{R}^3$, that is, there is a homeomorphism $\textbf{x}:\R^3\to M$ from $\R^3$ to $M$ --- the \textit{chart}. In the case of many parameterizations, the change of charts must smooth. Informally, the parametric manifold is a continuous deformation of the Euclidean space.
Figure~\ref{fig:geodesic} shows a schematic view of a $ 2 $-manifold.

Let $p$ be a point in the $3$-manifold $M$, and $\textbf{x}(x_1,x_2,x_3)$ be a parameterization of $M$. The \textit{tangent space} $T_pM$ in $p$ is the vector space generated by the vectors $\{\X{i}\textbf{x}_{|p}\}$ --- the derivatives of the coordinates curves of $\textbf{x}$ at $p$. As $T_pM$ admits a linear vector field structure, we can define a scalar product on each tangent space of $M$. If such procedure is ``smooth'' the result is a metric on $M$.

%===================================================
\subsection{Riemannian metric}\label{subsec:metric}\hfill\\ A \textit{Riemannian metric} in $M$ is a map $g$ that attributes to each point $p$ a scalar product $\dot{\cdot}{\cdot}_p$ in the tangent space $T_pM$, such that in coordinates,  $\textbf{x}(x_1,x_2,x_3)=p$, the function $g_{ij}(x_1,x_2,x_3):=\dot{\X{i}}{\X{j}}_p$ is smooth. This function does not depend on the coordinate system. We may use $g$ instead of $\dot{\cdot}{\cdot}$.

Expressing two tangent vectors $u$, $v$, at a point $p\in M$, in terms of the basis $\{\X{i}(p)\}$, $u=\sum u_i \X{i}(p)$ and $v=\sum v_i \X{i}(p)$, we obtain the scalar product:
\begin{equation}\label{eq:metric}
	\displaystyle\dot{u}{v}_p=\sum_{i,j=1}^3u_iv_j\dot{\X{i}}{\X{j}}(p).
\end{equation}
The metric tensor $g$ in $M$ is fully determined by the matrix $[g_{ij}]:=[\dot{\X{i}}{\X{j}}]$ and generalizes the classical Euclidean inner product of $\R^3$.
The pair $(M,g)$ is a \textit{Riemannian manifold}.
The chart $\textbf{x}$ bridges $M$ and $\R^3$, so that the \textit{pull-back} of the metric $g$ to $\R^3$ results in a Riemannian manifold $(\R^3,g)$ isometric to $(M,g)$.
This work investigates the design and visualization of Riemannian metrics on $\R^3$.

%===================================================
\subsection{Geodesics}\label{subsec:geodesics}\hfill\\ We now investigate the geodesics (rays) in the Riemannian manifold $(M, g)$. In the Euclidean metric, these curves have null acceleration (second derivative). Here, the analogous concept of ``acceleration'' is defined using the \textit{covariant derivative}. This is a way to calculate the derivative of tangent fields along curves in manifolds. 

A \textit{geodesic} in a Riemannian manifold $(M,g)$ is a curve $\gamma(t)=(x_1(t),x_2(t),x_3(t))$ with null \textit{covariant derivative}:
\begin{equation}\label{eq:geodesic_equation}\small
	\frac{D}{dt}\gamma'=\sum_{k=1}^{3}\left(x''_k+\sum_{i,j=1}^{n}\Gamma^k_{ij}x'_ix'_j\right)\frac{\partial}{\partial x^k}=0 \Longleftrightarrow x''_k+\sum_{i,j=1}^{3}\Gamma^k_{ij}x'_ix'_j=0, \,\, k=1,2,3.
\end{equation}
This differs from the classical by the addition of $\sum\Gamma^k_{ij}x'_ix'_j$, which includes the \textit{Christoffel symbols} $\Gamma^m_{ij}$  of $(M, g)$. 

To linearize System~\ref{eq:geodesic_equation}, we add new variables being the first derivatives $y_k = x'_k$, obtaining thus the \textit{geodesic flow} of $(M, g)$:
\begin{equation}\label{eq:geodesic_equation_bundle}
\left\{
\begin{array}{ll}
x'_k& = y_k  \\[0.0cm]
y'_k& =\displaystyle-\sum_{i,j=1}^{n}\Gamma^k_{ij}y_iy_j, \,\,\, k=1,2,\ldots,n. 
\end{array}
\right.
\end{equation}
To compute the geodesic flow in a Riemannian manifold we need its Christoffel symbols at each point. These can be computed in terms of the metric coefficients and its derivatives~\cite{carmo1992riemannian} by the formula:
\begin{equation}\label{eq:christofell_symbols}
	\displaystyle\Gamma^m_{ij}=\frac{1}{2}\sum_{k=1}^3\left( \X{i}g_{ik}+\X{j}g_{kj}-\X{k}g_{ij} \right)g^{km}.
\end{equation}
Where $[g^{ij}]$ is the inverse matrix of $[g_{ij}]$. 

%===================================================
\subsection{Exponential map}\label{s-exponential-map}\hfill\\
Let $p$ be a point in $M$ --- the \textit{observer}. A key idea in the ray tracing algorithm is to trace rays from $p$. This is also behind the \textit{exponential map}: a natural application encoding all the rays leaving $p$:
\begin{equation}
    \begin{array}{llll}
        \exp_p:& T_pM &\longrightarrow    & M \\
               & v          &\longmapsto& \gamma(\norm{v}). 
    \end{array}
\end{equation}
$\gamma$ is the geodesic satisfying $\gamma(0)=p$ and $\gamma'(0)=v$.
Intuitively, the exponential map takes a vector $v$ tangent at a point $p$, and runs a geodesic from $p$ towards the direction $v/\norm{v}$ until the parameter $\norm{v}$ (see Figure~\ref{fig:geodesic}).

A natural concern arises: Is $\exp_p$ well defined? The well-known \textit{Hopf--Rinow} theorem~\cite{carmo1992riemannian} gives us an answer. It states that $\exp_p$ is defined at the whole tangent space iff the Riemannian manifold is a complete metric space. In our case, $M$ diffeomorphic to $\R^3$, then $\exp_p$ is defined in the whole space. Additionally, the theorem say that every point in $M$ can be connected to $p$ by a geodesic. Thus, $\exp_p$ is surjective implying that every scene in $M$ can be ray traced using this map.

However, $\exp_p$ may not be injective: when there exists points in $M$ connected to $p$ by many geodesics. Visualizing such manifolds using \textit{rasterization} is infeasible.

The \textit{Hadamard} theorem~\cite{carmo1992riemannian} states that the exponential map will be injective if, additionally, $M$ has non-positive \textit{sectional curvature}. If we consider $M$ being the paraboloid, given by the graph of $f(x,y,z)=x^2+y^2+z^2$, the exponential map is not injective: there exists pairs of points in $M$ connected by many geodesics.
%===================================================

%===================================================
\section{Riemannian shading and illumination}\label{s-Riemannian_shading_illumination}
%===================================================
In computer graphics, \textit{Shading} is the process of assigning a color to a pixel. Classic approaches to perform such tasks are not suited for Riemannian manifolds due to the nonlinear nature of their rays.

%===================================================
\subsection*{Riemannian shading}\hfill\vspace{0.05cm}\\ Consider a viewing $2$-sphere $\S^2_p$ centered in an \textit{observer} point $p$ in a $3$-manifold $M$. We give a color for each \textit{ray direction} in the the observer field $V\subset\S^2_p$ by tracing a ray; $\S^2_p \cap V$ carries the image. We call this procedure \textit{Riemannian shading}.
Specifically, the sphere $\S^2_p$ is centered at the origin of $T_pM$. For each direction $v \in V\subset\S^2_p$, we attribute a color $c$ by launching a ray $\gamma(t)$ from $p$ towards $v$ using the exponential map $exp_p$. If $\gamma$ intersects a scene object, we define an RGB color. This is the \textit{Riemannian shading} $c:\S^2_p\to \mathcal{C}$, where $\mathcal{C}$ is a color space.

%===================================================
\subsection*{Riemannian illumination}\hfill\vspace{0.05cm}\\
The \textit{Riemannian illumination} of a point $q$ in an embedded surface $S\subset M$ comes from direct geodesics connecting $q$ to the light sources and indirect geodesics connecting other Riemannian-illuminated points to $q$. The \textit{radiant intensity} at $q$ is modeled using the \textit{Lambertian reflectance}, which depends on the Riemannian metric, or more generally from a BRDF. 
For a point $p$ in $M$, we compute the Riemannian shading using ray tracing and Riemannian~illumination.

Classical ray tracing~\cite{whitted1979improved} approximates physical illumination. Our Riemannian ray tracing model can be also used to compute a Riemannian shading function for local or global illumination.
This work uses pseudo-color based on properties of the space, such as a coordinated point, to define the Riemannian shading~function.

%===================================================
\subsection{Ray Marching}\label{subsec:numerical_integration}\hfill\\
To launch a ray we need to solve the geodesic flow (Eq~\ref{eq:geodesic_equation_bundle}) and subsequently compute the intersection of the given ray with the scene objects.
In general, our rays will be {\em curved} and are solutions of the geodesic flow which we have resorted to numerical integration methods.

We use Euler's method for integration in the parameterization image, since there we use the Euclidean metric.
The Euler's numerical integration method approximates a ray $\gamma$ starting at $p$ in the direction $v$ by a polygonal $\{p_i\}$:
\begin{equation}\label{eq:geodesic_euler}
\left\{
\begin{array}{l}
p_{i+1}=p_{i} + h\cdot \widetilde{\gamma}'(0)\\
v_{i+1}=v_{i} + h\cdot \widetilde{\gamma}''(0)
\end{array}
\right.
\end{equation}
where $h$ is the integration step and $\widetilde{\gamma}(t)$ is the ray satisfying $\widetilde{\gamma}(0)=p_i$ and $\widetilde{\gamma}'(0)=v_i$. We use Equation~\ref{eq:geodesic_equation_bundle} to compute $\widetilde{\gamma}''(0)$. 

The polygonal $\{p_i\}$ is used to ray trace a scene in $(M,g)$. When $h\to \epsilon$ the scene is accurately rendered.
For the ray-scene intersection, we check the for each segment given by the polygonal approximation given by Equation~\ref{eq:geodesic_euler} (i.e., ray marching).

For now, we are testing the intersection with each segment sequentially. In future works, we pretend to use many segments of the polygonal $\{p_i\}$ to increase the GPU occupancy and optimize the intersection computations. This idea was suggested by Ingo Wald during GTC 2020.

We could use the Runge–Kutta method to approximate solutions of the geodesic flow. Instead, we consider the Euler method since it uses fewer computations giving GPU performance.
%===================================================

%===================================================
\section{Examples of Riemannian metrics}\label{s:examples-manifolds}
%===================================================
This section presents two examples of Riemannian metrics in $ \R^3 $. The first is given by pushing the metric from the graph of three-dimensional functions, the other comes from deformations of $ \R^3 $  given by diffeomorphisms.

%===================================================
\subsection{Graph of a function}\label{ss-graph}\hfill\\
Let $f:\R^3\to \R$ be a smooth function. We construct a Riemannian metric $g$ in $\R^3$ which is the pullback of the metric of the \textit{graph of $f$}:
\begin{equation*}\label{eq:graph_f}
   M_f=\{(x_1,x_2,x_3,x_4)\in\R^4|\,\,f(x_1,x_2,x_3)=x_4\}.
\end{equation*}

$M_f$ is parameterized in the obvious way by $\textbf{x}(x_1,x_2,x_3)=(x_1,x_2,x_3,f(x_1,x_2,x_3))$. The \textit{tangent space} $T_pM_f$ at a point $p$ is generated by the vectors:
\begin{equation}\label{eq:partials}
\X{i}(p)=(e_i,f_i(p)). 
\end{equation}
Where $f_i(p)$ is the partial derivative of $f$ in the standard direction $e_i$. We use the Euclidean metric of $\R^4$ to induce a Riemannian metric $g$ in $M_f$.

Let $p$ be a point in $M_f$, we compute the metric $g$ at $p$. Replacing the expression of Equation~\ref{eq:partials} in $\dot{\X{i}}{\X{j}}$, we obtain the coefficients of the Riemannian metric:
\begin{equation}\label{eq-metric-graph}
g_{ii}=1+f_i^2\,\, \mbox{and } g_{ij}=f_if_j,\,\,\mbox{ if } i\neq j,
\end{equation}
which are smooth functions of $M_f$, then $g$ is \textit{Riemannian}. In terms of measure, the metric $g$ makes the volume vary beautifully because its \textit{volume formula}, used to compute integrals in $ (M_f, g) $, has the form $ \det [g_ {ij}] = 1+ \norm {\nabla f} ^ 2 $. Then, it only matches with the standard if $ \nabla f = 0 $.

%%%%%%geodesics
We present the geodesics of $(M_f,g)$. A direct calculation using the software \textit{Maple} implies in $g^{ii}=\frac{f_j+f_k}{1+\norm{\nabla f}^2}$, with $i$, $j$, $k$ being all distinct, and $g^{ij}=-\frac{f_if_j}{1+\norm{\nabla f}^2}$, if $ i\neq j$.
Substituting these in Equation~\ref{eq:christofell_symbols}, and using software \textit{Maple}, we obtain a simple formula for the Christoffel symbols of $(M_f, g)$:
\begin{equation}\label{eq:christofell_symbols_graph}
   \displaystyle\Gamma^m_{ij}=\frac{f_mf_{ij}}{1+\norm{\nabla f}^2}.
\end{equation}
Where $f_{ij}$ is the second order partial derivative in the direction $e_i$, and $e_j$. 
Replacing the Christoffel symbols of $M_f$ in this Equation~\ref{eq:christofell_symbols} we obtain.
\begin{equation}\label{eq:geodesic_equation_bundle_graph}
    \left\{
    \begin{array}{ll}
         x'_k& = y_k  \\
          y'_k& =\displaystyle-\sum_{i,j=1}^{3}\frac{f_kf_{ij}}{1+\norm{\nabla f}^2}y_iy_j, \,\,\, k=1,2,3. 
    \end{array}
    \right.
\end{equation}

\begin{figure}[hh]
    \centering
    \begin{tabular}{cc}
        \includegraphics[width=0.48\columnwidth]{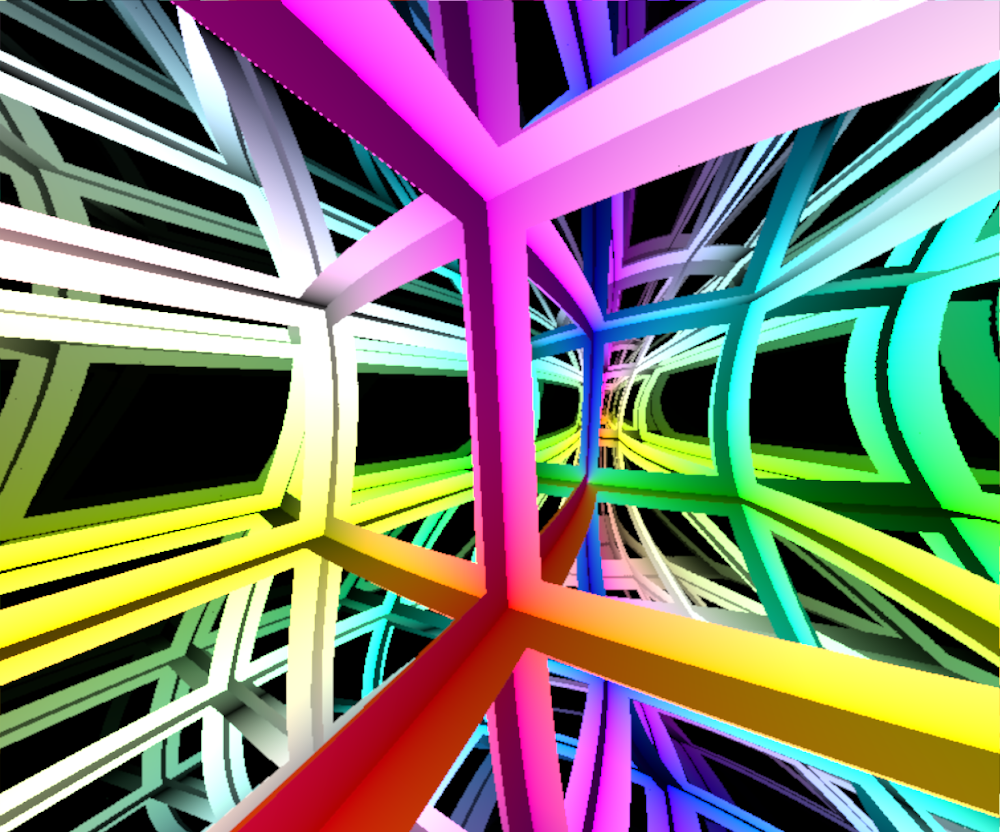}
        &
        \includegraphics[width=0.48\columnwidth]{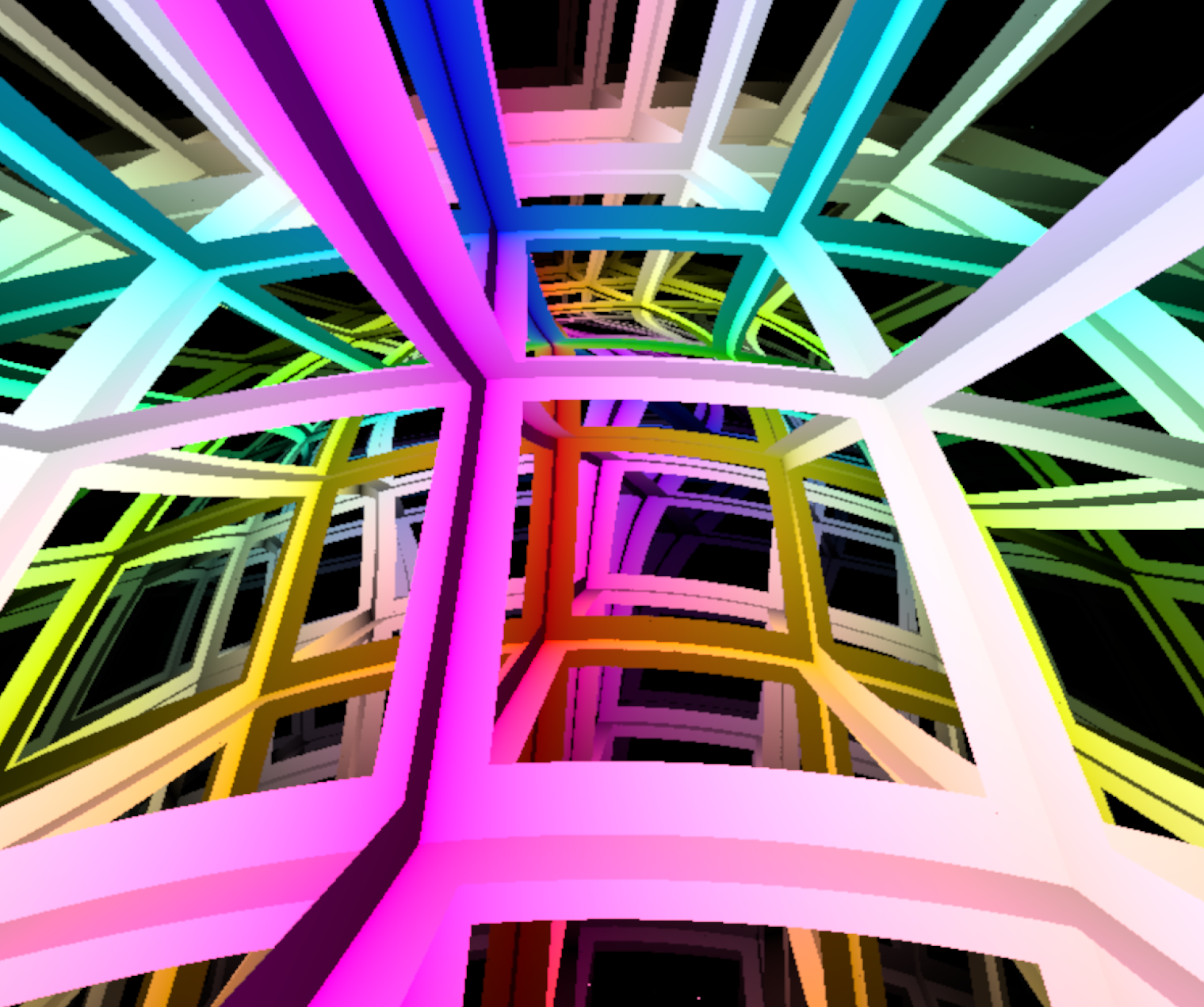}
    \end{tabular}
    \caption{The space $\R^3$ endowed with the geometry of the graph of the function $f(x,y,z)=x^2+y^2-z^2$. On the left, a look towards direction $x$. On the right, a look in the direction $z$.}
    \label{fig:graph}
\end{figure}

\vspace{0.3cm}
For a concrete example, consider $M_f$ being the graph of the polynomial function $f(x,y,z)=x^2+y^2-z^2$.
The Christoffel symbols of this parameterization are computed using Equation~\ref{eq:christofell_symbols_graph}. Replacing the result in Equation~\ref{eq:geodesic_equation_bundle_graph} we obtain a geodesic flow. 
Figure~\ref{fig:graph} gives an inside view of the space $\R^3$ endowed with the geometry pulled back from $M_f$ by the parameterization $\Phi(x,y,z)=(x,y,z,f(x,y,z))$. In the left image, we positioned the camera looking towards direction $x$. On the right image, the camera points at the direction $z$.
The scene is composed of a regular grid in $\R^3$ distorted by $\Phi$.

%===================================================
\subsection{Diffeomorphisms}\label{ss-diffeo}\hfill\\
Let $\Phi:\R^3\to \R^3$, given by $\Phi(p)=(x_1(p), x_2(p), x_3(p))$, be a \textit{diffeomorphism}. We put a metric in $\R^3$ based in the deformations provided by $\Phi$. Our base manifold in this case is $\R^3$ parameterized by $\Phi$, and its \textit{associated base} is given by:
\begin{equation}\label{eq:diffeo_partials}
\X{i}=(\X{i}x_1, \X{i}x_2, \X{i}x_3). 
\end{equation}
We \textit{pullback} the Euclidean metric of $\R^3$ through the differential of $\Phi$, this will provide a new metric $g$ on $\R^3$.

The metric $g$ at a point $p\in\R^3$ is completely determined by the \textit{metric tensor} $[g_{ij}]:=[\dot{\X{i}}{\X{j}}]$. Thus subtituing Equation~\ref{eq:diffeo_partials} in $\dot{\X{i}}{\X{j}}$, we obtain the following formula:
\begin{equation}\label{eq:diffeo-metric}
g_{ij}=\displaystyle\sum_{k=1}^3\X{i}x_k\X{j}x_k. 
\end{equation}
As $g_{ij}$ are smooth functions, $g$ is \textit{Riemannian}.

%geodesics
We now compute the geodesics of $(\R^3,g)$. By Equation~\ref{eq:geodesic_equation} we must first calculate the Christoffel symbols. Replacing the metric in Equation~\ref{eq:diffeo-metric} in Christoffel symbols Equation~\ref{eq:christofell_symbols} and using software \textit{Maple}, we obtain a simple formula:
\begin{equation*}
   \displaystyle\Gamma^m_{ij}=\sum_{s=1}^3\left(\frac{\partial^2}{\partial x_i\partial x_j}x_s\cdot(-1)^{s+m}\cdot\frac{\det J_{sm}}{\det J}\right).
\end{equation*}
Where $J$ is the \textit{Jacobian matrix} of $\Phi$, and $J_{sm}$ is $J$ without the $s$-line and the $m$-column: the $(s,m)$-minor of $J$.
The term $(-1)^{s+m}\cdot\frac{\det J_{sm}}{\det J}$ in the equation is the $(m,s)$-coefficient of the inverse of $J$. Thus we obtain a beautiful formula.

\begin{theorem}\label{th:Christoffel-symbols}
	Let $\Phi(p)=(x_1(p), x_2(p), x_3(p))$ be a diffeomorphism of $\R^3$. Then, the Christoffel symbols of $\Phi$ are given by:
\begin{equation}\label{eq:christofell_symbols_diffeo}
   \displaystyle\Gamma^m_{ij}=\sum_{s=1}^3\left(\frac{\partial^2}{\partial x_i\partial x_j}x_s\cdot[J^{-1}]_{ms}\right).
\end{equation}
\end{theorem}

Theorem~\ref{th:Christoffel-symbols} says that the Christoffel symbols of $(\R^3,g)$ can be computed using only the Jacobian and Hessian operators of the parameterization $\Phi$. For the best of our knowledge, this is the first time such a formula appears. Equation~\ref{eq:christofell_symbols_diffeo} will be very useful when the parameterization $\Phi$ be a composition of parameterizations.
 
Replacing the Christoffel symbols in the Eq.~\ref{eq:geodesic_equation}, we obtain its geodesic flow:
\begin{equation}\label{eq:geodesic_equation_bundle_diffeo}
    \left\{
    \begin{array}{ll}
         x'_k& = y_k  \\
          y'_k& =\displaystyle-\sum_{i,j,s=1}^{3}\left(\frac{\partial^2}{\partial x_i\partial x_j}x_s\cdot[J^{-1}]_{ms}\right)y_iy_j, \,\,\, k=1,2,3. 
    \end{array}
    \right.
\end{equation}

\begin{figure}[hh]
    \centering
    \includegraphics[width=1\columnwidth]{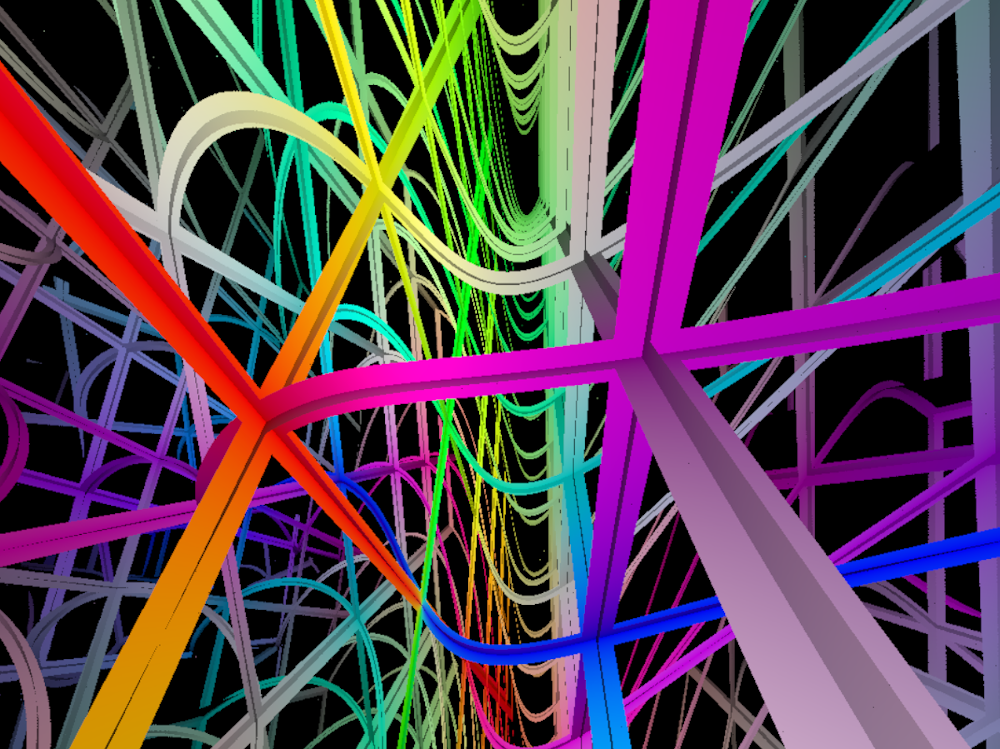}
    \caption{The space $\R^3$ deformed using a ``twist'' along the planes perpendicular to the $z$-axis. The RGB colors are given by the $(x,y,z)$ coordinates of the hit points.}
    \label{fig-twisted-space}
\end{figure}

\vspace{0.5cm}
For an explicit example of a deformation of $\R^3$, we consider the following twist~map
\begin{equation*}
\Phi(x,y,z)=(\cos(z)x-\sin(z)y, \sin(z)x+\cos(z)y ,z).
\end{equation*}
The Christoffel symbols of this parameterization are computed using Equation~\ref{eq:christofell_symbols_diffeo}. Replacing the result in Equation~\ref{eq:geodesic_equation_bundle_diffeo} we obtain a geodesic flow. We give an inside view of this ``twisted'' $\R^3$ by tracing rays using Riemannian shading. Figure~\ref{fig-twisted-space} shows an immersive view of a regular grid in $\R^3$ distorted by $\Phi$.
%===================================================

%===================================================
\section{Deforming the space}
%===================================================
This section describes two convenient ways of accumulating local deformations of $\R^3$. The first is an extension of Subsection~\ref{ss-graph} consisting of summing functions. The second deals with the composition of deformations. In both cases, the functions and deformations are modeled using \textit{Gaussians}.
Many techniques in Computer Graphics depends on these functions due to their flexibility in modeling and smoothness. 

%===================================================
\subsection{Local metric deformations}\hfill\\
We model graphs and deformations using the three-dimensional \textit{Gaussian function}
\begin{equation}
f(x,y,z)=a\cdot\exp\left(-\frac{(x-x_0)^2}{2\sigma_x^2}-\frac{(y-y_0)^2}{2\sigma_y^2}-\frac{(z-z_0)^2}{2\sigma_z^2}\right).
\end{equation}
Where $a$ is the amplitude, $(x_0,y_0,z_0)$ is the center, and $\sigma_x$, $\sigma_y$, $\sigma_z$ are the spreads.

\subsection*{Using graphs}
\begin{figure}[hh]
    \centering
    \begin{tabular}{cc}
        \includegraphics[width=0.32\columnwidth]{figures/no-deform.png}
        &
        \includegraphics[width=0.32\columnwidth]{figures/1-local-deform1.png}\\
        \includegraphics[width=0.32\columnwidth]{figures/deform.png}
        &
        \includegraphics[width=0.32\columnwidth]{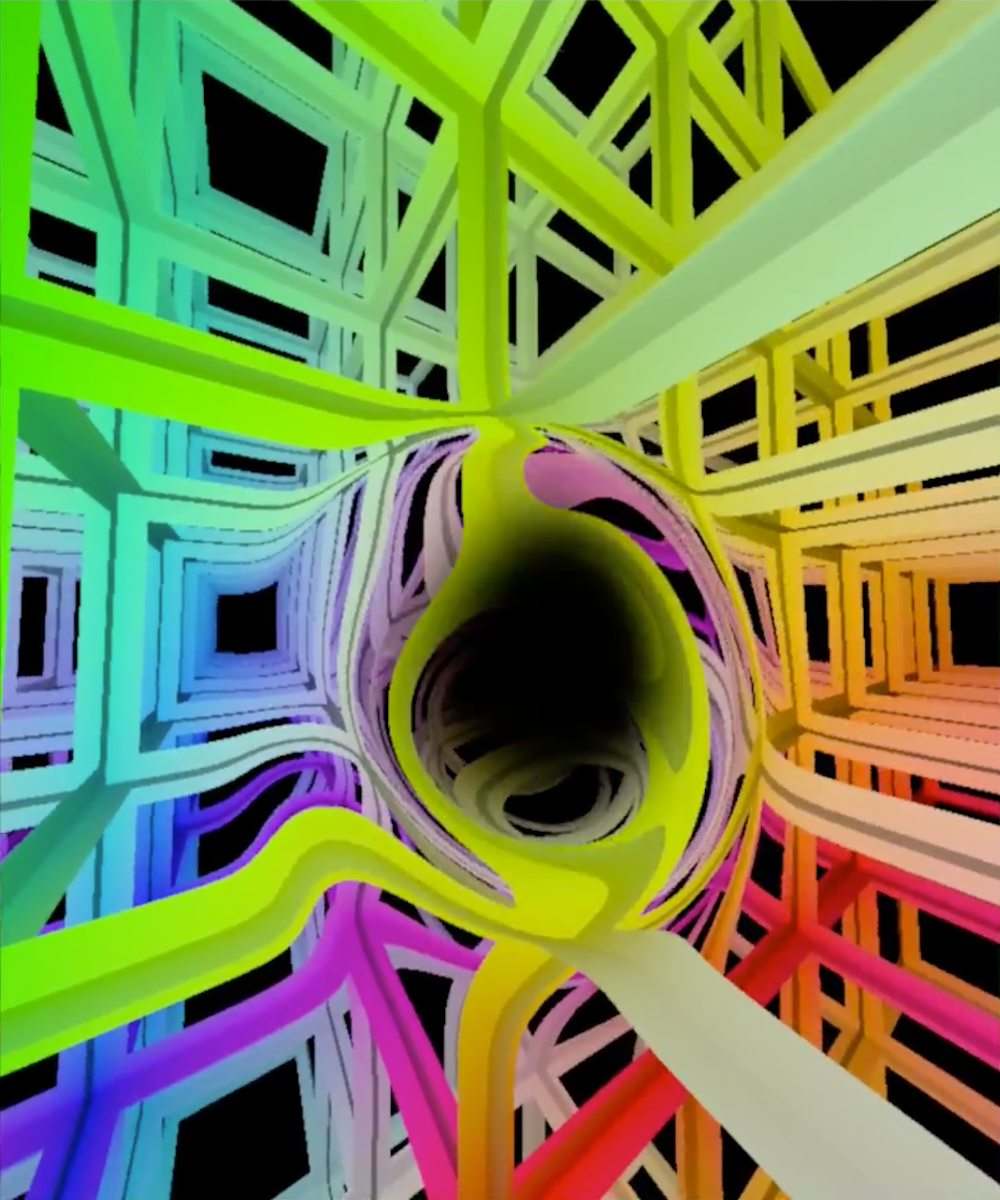}
    \end{tabular}
    \vspace{-0.2cm}
    \caption{The space $\R^3$ endowed with no deformation, local deformation, and global deformation.}
    \label{fig-local-graph-deformation}
\end{figure}

\hfill\\
Using Subsection~\ref{ss-graph}, we visualize the graph $M_f=\{p\in\R^4|z=f(x,y,z)\}$. Figure~\ref{fig-local-graph-deformation} shows inside views of $M_f$ by varying the parameters of $f$. The first image presents the case of zero amplitude. The second image illustrates a small deformation given by a small amplitude and spreads values. The third and fourth images have small amplitude with lager spreads, and vice-versa. In all the cases, the scene is composed of a regular grid and the ray tracing has a small (black) fog component. This is why in the fourth image there is a black region in the center of the deformation.

\subsection*{Using diffeomorphism}\hfill\\
We present a formula to deform a neighborhood of a point $q$ in $\R^3$. Consider a Gaussian function $f$ centered at $q$, choosing small spreads values for $f$, the formula
\begin{eqnarray}\label{eq-local-diffeo}
\Phi(p)=p+f(p) \cdot v 
\end{eqnarray}
deforms the neighborhood of $q$ at the direction $v$.

For a more refined formula, replace $v$ by a vector field $V$. The resulted formula $\Phi(p)=p+f(p) \cdot V(p)$ deforms a neighborhood of $q$ following the instructions of $V$.

%===================================================
\subsection{Accumulating  deformations}\hfill\\
We present two ways of accumulating deformations: summing the functions or the deformations presented above, and composing diffeomorphism.

\subsection*{Summing functions}\hfill\\
Let $f_1,f_2,\ldots,f_k$ be functions taking values in $\R^3$. The graph of the function $f=f_1+f_2+\ldots+f_k$ is given by
\begin{equation*}
M_f=\{p\in\R^4|(f_1+f_2+\ldots+f_k)(x_1,x_2,x_3)=x_4\}.
\end{equation*}

As we saw in Subsection~\ref{ss-graph}, the manifold $M_f$ endows the metric of $\R^4$. Equation~\ref{eq-metric-graph} says that this metric is given by $g_{ij}=1+\left(\frac{\partial f}{\partial x_i}\right)^2$ and $g_{ij}=\frac{\partial f}{\partial x_i}\frac{\partial f}{\partial x_j}$ if $i\neq j$. Then the metric $[g_{ij}]$ is easily computed because 
\begin{equation*}
\frac{\partial f}{\partial x_i}=\frac{\partial (f_1+f_2+\ldots+f_k)}{\partial x_i}=\frac{\partial f_1}{\partial x_i}+\frac{\partial f_2}{\partial x_i}+\ldots+\frac{\partial f_k}{\partial x_i}.
\end{equation*}
In other words, to compute the metric $g$ we only need to store the partial derivatives of each function $f_i$. The pair $(M_f, g)$ is the desired Riemannian manifold.

We now compute the geodesic flow of $(M_f, g)$. The geodesic Equation~\ref{eq:geodesic_equation} is given by the Christoffel symbols of $(M_f, g)$, which are presented as $\Gamma^m_{ij}=\frac{f_mf_{ij}}{1+\norm{\nabla f}^2}$ by  Eq.~\ref{eq:christofell_symbols_graph} in terms of $f$.
Therefore, the symbols $\Gamma^m_{ij}$ can be computed using the first and second partial derivatives of the functions $f_i$'s.

\begin{figure}[hh]
    \centering
    \begin{tabular}{ccc}
        \includegraphics[width=0.31\columnwidth]{figures/no-deform.png}
        &
        \includegraphics[width=0.31\columnwidth]{figures/1-local-deform1.png}
        &
        \includegraphics[width=0.31\columnwidth]{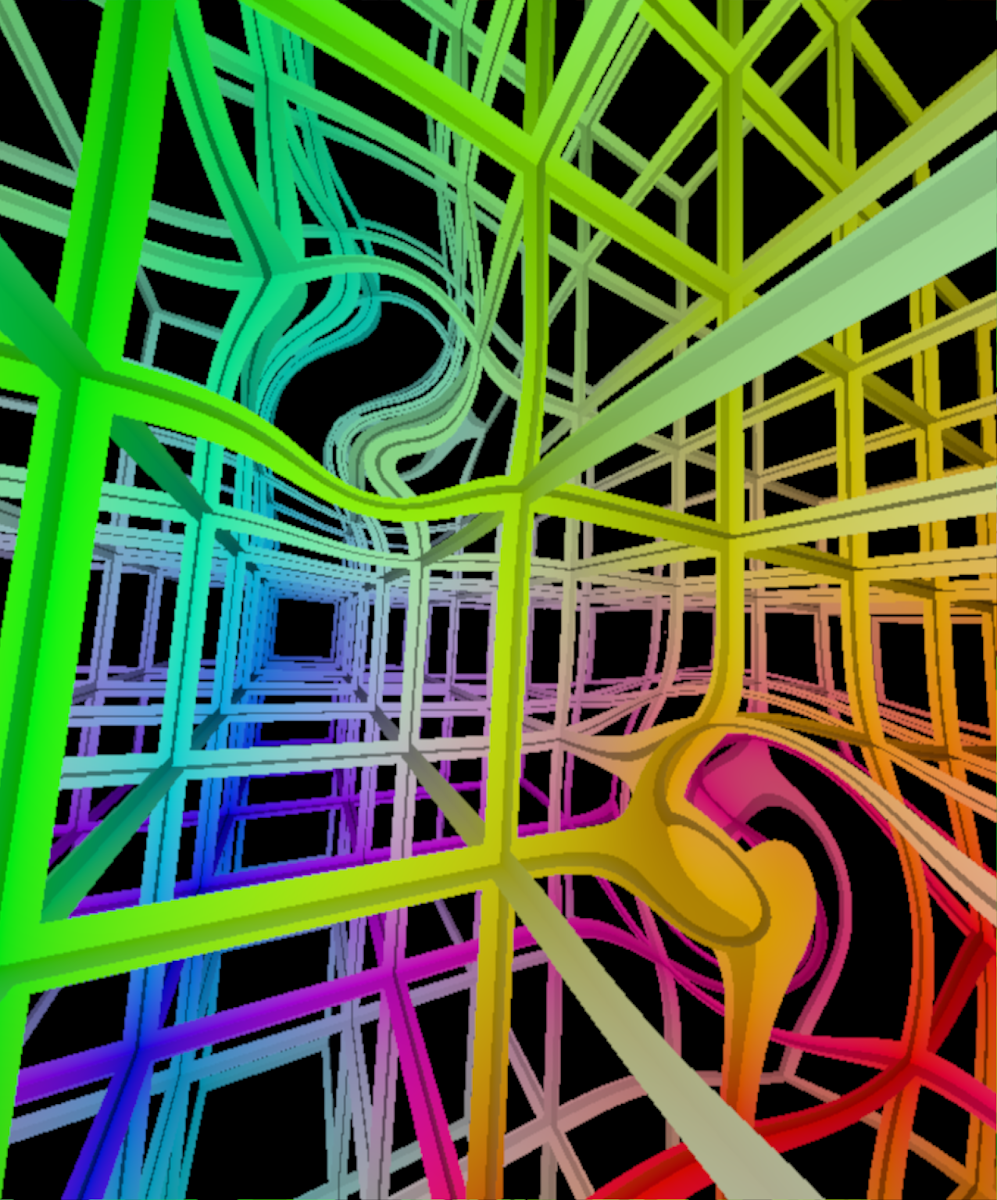}
    \end{tabular}
    \caption{Space $\R^3$ endowed with no deformation, one local deformation, and two local deformations.}
    \label{fig-sum-of-functions}
\end{figure}

To show the effects in visualizing graphs of summed functions, consider a regular grid embedded in $\R^3$ (Left image in Figure~\ref{fig-sum-of-functions}). Adding a unique Gaussian function in the summation (middle image in Figure~\ref{fig-sum-of-functions}). Summing two Gaussian functions (right image in Figure~\ref{fig-sum-of-functions}).

\subsection*{Composing diffeomorphisms}\hfill\\
Let $\Phi_1,\Phi_2,\ldots,\Phi_k$ be a sequence of diffeomorphisms of $\R^3$, we pretend to compute the Riemannian metric $g$ on $\R^3$ that carries the deformations of the composition $\Phi=\Phi_1\circ\Phi_2\circ\ldots\circ\Phi_k$. Then we derive the geodesic flow of the resulting Riemannian manifold $(\R^3,g)$ for a possible ray tracing.

Equation~\ref{eq:diffeo-metric} states that the metric $g$ is given by $g_{ij}=\sum_k\X{i}x_k\X{j}x_k$, where $\Phi(p)=(x_1(p),x_2(p),x_3(p))$. In other words, to compute the metric coefficients $g_{ij}$, we only need the Jacobian matrix $J(\Phi)$ of $\Phi$ which is easily obtained using the \textit{chain rule} formula $J(\Phi)=J(\Phi_1)\cdot J(\Phi_2)\cdot\ldots \cdot J(\Phi_k)$. 

To compute the geodesic flow of the Riemannian manifold $(\R^3,g)$, we need its Christoffel symbols, which are provided by the Jacobian $J(\Phi)$ and the Hessian $H(\Phi)$ operator of $\Phi$ (from Theorem~\ref{th:Christoffel-symbols}) using the formula
$  \Gamma^m_{ij}=\sum_s\left(\frac{\partial^2}{\partial x_i\partial x_j}x_s\cdot[J^{-1}]_{ms}\right)$.
Specifically, to compute $\Gamma^m_{ij}$, we invert the Jacobian $J(\Phi)$ of $\Phi$ using the formula
\begin{equation*}
J^{-1}(\Phi)=J^{-1}(\Phi_k)\cdot J^{-1}(\Phi_{k-1})\cdot\ldots \cdot J^{-1}(\Phi_1).
\end{equation*}
It is not direct to compute the Hessian component $\frac{\partial^2}{\partial x_i\partial x_j}x_s$ of $\Phi$, where $x_s(p)$ is the $s$-coordinate of $\Phi(p)$. Firstly, we consider the composition of only two diffeomorphisms $\Phi=\Phi_1\circ\Phi_2$, then we extend it to the composition of diffeomorphisms.

The Hessian matrix of the coordinate of $\Phi=\Phi_1\circ\Phi_2=(x_1\circ\Phi_2,x_2\circ\Phi_2,x_3\circ\Phi_2)$, that is $H(x_s\circ\Phi_2)$, is given by 
\begin{equation}\label{eq-hessian-composition}
H(x_s\circ\Phi_2)_p=J^T(\Phi_2)_p\cdot H(x_s)_{\Phi_2(p)}\cdot J(\Phi_2)_p + \nabla x_s(\Phi_2(p))\cdot H(\Phi_2)_p
\end{equation}

We extended Equation~\ref{eq-hessian-composition} to $x_s\circ\Phi_2\circ\Phi_3$, that is, we consider the composition of three diffeomorphisms $\Phi=\Phi_1\circ\Phi_2\circ\Phi_3$. For this, we apply Equation~\ref{eq-hessian-composition} to $H((x_s\circ\Phi_2)\circ\Phi_3)_p$, then after some computations we obtain:
{\small
\begin{equation*}
\begin{array}{lll}
H(x_s\circ\Phi_2\circ\Phi_3)_p&=&J^T(\Phi_2\circ\Phi_3)_p\cdot H(x_s)_{\Phi_2\circ\Phi_3(p)}\cdot J(\Phi_2\circ\Phi_3)_p\\
 &+& J^T(\Phi_3)_p\cdot\nabla x_s(\Phi_2\circ\Phi_3(p))\cdot H(\Phi_2)_{\Phi_3(p)}\cdot J(\Phi_3)_p \\
 &+& \nabla (x_s\circ\Phi_2)(\Phi_3(p))\cdot H(\Phi_3)_p
\end{array}
\end{equation*}
}

The general formula is obtained through an inductive argument. We simplify the notation by considering $\Phi_{[i,j]}:=\Phi_i\circ\ldots\circ\Phi_j$, for each $i<j$. Therefore 
{\small
\begin{align*}
H(x_s\circ\Phi_{[2,k]})_p&=J^T(\Phi_{[2,k]})_p\cdot H(x_s)_{\Phi_{[2,k]}(p)}\cdot J(\Phi_{[2,k]})_p\\
&+\displaystyle \sum_{i=3}^kJ^T(\Phi_{[i,k]})_p\cdot\nabla 
(x_s\circ\Phi_{[2,i-2]})(\Phi_{[i-1,k]}(p))\cdot H(\Phi_{i-1})_{\Phi_{[i,k]}(p)}\cdot J(\Phi_{[i,k]})_p\\
&+\nabla(x_s\circ\Phi_{[2,k-1]})(\Phi_{k}(p))\cdot H(\Phi_{k})_p
\end{align*}
}
From a computational point of view, this is an important formula because it computes $H(x_s\circ\Phi_{[2,k]})_p$ in terms of the Jacobians and Hessians of each diffeomorphism $\Phi_i$ separately. Therefore, during the computations only the partial derivatives of each diffeomorphism must be stored, the partial derivatives of the compositions are derived from them.
%===================================================
% \subsection{Time/User--dependent metrics}\hfill\\

% . Exploring especial effects
%===================================================

%===================================================
% \section{Efficient storing of deformations using hierarchical structures}
% %===================================================
% . For graphs

% . For diffeomorphisms 
%===================================================

\newpage

\bibliographystyle{amsplain}
\bibliography{ray3d}

\end{document}